\documentclass[
amsmath,amssymb,11pt,onecolumn,aps,prx,floatfix]{revtex4-2}

\usepackage[english]{babel}
\usepackage{multirow}
\usepackage[thinspace]{SIunits}
\usepackage{xspace}
\usepackage{ulem}
\usepackage{soul}
\usepackage{graphicx}
\graphicspath{{figures//}} 
\usepackage{dcolumn}
\usepackage{bm}
\usepackage{color}
\usepackage{wrapfig}
\usepackage{hyperref}
\usepackage{array}
\usepackage{booktabs}
\usepackage{multirow}
\usepackage{amssymb}

\usepackage{marvosym}
\hypersetup{
    unicode=false,          
    pdftoolbar=true,        
    pdfmenubar=true,        
    pdffitwindow=false,     
    pdfstartview={FitH},    
    pdftitle={My title},    
    pdfauthor={Author},     
    pdfsubject={Subject},   
    pdfcreator={Creator},   
    pdfproducer={Producer}, 
    pdfkeywords={keyword1} {key2} {key3}, 
    pdfnewwindow=true,      
    colorlinks=true,       
    linkcolor=blue,          
    citecolor=blue,        
    filecolor=blue,      
    urlcolor=blue,       
    plainpages=false        
}

\newcommand{\LNO}{\mbox{La$_3$Ni$_2$O$_7$}\,\xspace}
\newcommand{\Alg}{$A_{\rm {1g}}$\,\xspace}

\newcommand{\Blg}{${B_{1g}}$\xspace}
\newcommand{\BZg}{${B_{2g}}$\xspace}

\newcommand{\wn}{\rm{cm}$^{-1}$\,}

\usepackage{xparse}

\makeatletter
\AtBeginDocument{
    \setcounter{secnumdepth}{2}

    \renewcommand{\thefigure}{\arabic{figure}}
    \renewcommand{\thetable}{\Roman{table}}

    \renewcommand{\fnum@figure}{\normalfont FIG.~\thefigure}
    \renewcommand{\fnum@table}{\normalfont TABLE~\thetable}

    \setlength{\parindent}{1em}

    \renewcommand\section{\@startsection
        {section}{1}{0pt}
        {20pt plus 3pt minus 2pt}
        {10pt}
        {\normalfont\normalsize\bfseries\centering}}

    \renewcommand\subsection{\@startsection
        {subsection}{2}{0pt}
        {12pt plus 3pt minus 2pt}
        {6pt}
        {\normalfont\normalsize\bfseries\centering}}

    \def\@hangfrom@section#1#2{\@hangfrom{#1}#2}
    \def\@hangfroms@section#1#2{\@hangfrom{#1}#2}

    \let\originalsection\section

    \RenewDocumentCommand{\section}{s o m}{
        \IfBooleanTF{#1}
            {\originalsection*{\MakeUppercase{#3}}}
            {\IfNoValueTF{#2}
                {\originalsection{\MakeUppercase{#3}}}
                {\originalsection[#2]{\MakeUppercase{#3}}}}
    }
}
\makeatother

\begin{document}

\def\scititle{Competing magnetism and weakly anisotropic superconductivity in pressurized single-crystal \LNO}
\title{\bfseries \boldmath \scititle}

\author{Hengyuan Zhang$^{1,*}$,
Deyuan Hu$^{1,*}$,
Jun Shu$^{2,3,*}$,
Junfeng Chen$^{1}$,
Lauro B. Braz$^{4,5}$,
Dengjing Wang$^{3}$,
Chaoxin Huang$^{1}$,
Bing Shen$^{1}$,
Ilya M. Eremin$^{4}$,
Hualei Sun$^{6,\textrm{\Letter}}$,
Ge He$^{2,7,\textrm{\Letter}}$,
Jun Shen$^{2,\textrm{\Letter}}$ and
Meng Wang$^{1,\textrm{\Letter}}$}

\affiliation{
$^1$ Institute of Neutron Science and Technology\mbox{,} Guangdong Provincial Key Laboratory of Magnetoelectric Physics and Devices\mbox{,} School of Physics\mbox{,} Sun Yat-Sen University\mbox{,} Guangzhou\mbox{,} Guangdong 510275\mbox{,} China\\
$^2$ School of Mechanical Engineering\mbox{,} Beijing Institute of Technology\mbox{,} Beijing 100081\mbox{,} China \\
$^3$ Department of Applied Physics\mbox{,} Wuhan University of Science and Technology\mbox{,} Wuhan 430081\mbox{,} China\\
$^4$ Institut für Theoretische Physik III, Ruhr-Universität Bochum, D-44801 Bochum, Germany.\\
$^5$ Instituto de Fısica, Universidade de Sao Paulo, Rua do Matao 1371, Sao Paulo, Sao Paulo 05508-090, Brazil.\\
$^6$ School of Science, Sun Yat-Sen University, Shenzhen, 518107, China\\
$^7$ Beijing Key Laboratory of Quantum Matter State Control and Ultra-Precision Measurement Technology\mbox{,} Beijing Institute of Technology\mbox{,} Beijing 100081\mbox{,} China\\
$^{*}$ These authors contributed equally: Hengyuan Zhang, Deyuan Hu and Jun Shu\\
\textrm{\Letter} e-mail: 
sunhlei@mail.sysu.edu.cn;
ge.he@bit.edu.cn;
jshen@bit.edu.cn;
wangmeng5@mail.sysu.edu.cn
}

\begin{abstract}
\section*{Abstract}
Unconventional high-temperature superconductivity often emerges as magnetic order is destabilized, but whether the recently discovered bilayer nickelates follow this organizing principle remains unresolved. Here we use polarization-resolved electronic Raman spectroscopy to map the pressure evolution of magnetism and superconductivity in single-crystalline La$_3$Ni$_2$O$_7$. In addition to the known spin-density wave, we uncover a field-sensitive collective mode below approximately 130~K between 9 and 14~GPa. Its symmetry and pressure evolution point to a competing double-$Q$ magnetic/charge instability that strengthens on approaching the pressure-induced structural transition, but disappears as superconductivity develops. In the superconducting regime, symmetry-selective pair-breaking excitations provide direct determination of the gap structure in pressurized bulk La$_3$Ni$_2$O$_7$. A simultaneous analysis of the \Blg and \BZg responses reveals nodeless, weakly anisotropic multiband $s$-wave gaps on the resolved Fermi-surface pockets and disfavors the proposed $d$-wave state. By connecting the pressure-driven competition and suppression of magnetic orders with the emergence and gap symmetry of superconductivity, our results identify fluctuations among competing magnetic states as a plausible ingredient for pairing. Taken together, these findings establish bilayer nickelates as a new platform for disentangling universal links between collective electronic fluctuations and unconventional pairing from material-specific forms of order across correlated superconductors.
\end{abstract}

\maketitle

\section{Introduction}

The discovery of superconductivity with transition temperatures approaching 80 K in pressurized bilayer nickelates has established \LNO as a new platform for unconventional high-temperature superconductivity beyond cuprates and iron-based superconductors \cite{Sun:2023,Zhang:2024,WangNN:2024}. In contrast to infinite-layer nickelates, \LNO contains coupled NiO$_2$ bilayers with active Ni $d_{x^2-y^2}$ and $d_{z^2}$ orbitals, strong hybridization with apical oxygen states, and substantial interlayer electronic coupling \cite{WangReview:2024,Qu:2024,Bejas:2026,Gu:2025,ZhangNC:2024,YangYF:2023}. These ingredients place bilayer nickelates in a distinct multiorbital and multilayer regime, where electronic correlations, orbital selectivity and interlayer exchange may all participate in pairing. Determining the superconducting gap structure, and identifying the parent electronic phase from which superconductivity emerges, are therefore central questions for understanding the pairing mechanism in nicklates.

The pairing symmetry of bilayer nickelates remains unsettled. Theoretical studies have proposed competing sign-changing $s$-wave and $d$-wave states, which can be close in energy and sensitive to pressure, lattice structure, orbital splitting and Fermi-surface topology \cite{WangReview:2024,Liu:2023,Ilya:2024,Fan:2026,Nie:2026}. Experimentally, important progress has been made in superconducting thin films, where scanning tunnelling microscopy and spectroscopy (STM/STS) and angle-resolved photoemission spectroscopy (ARPES) have reported nodeless gaps, multiple gap scales and electron-boson coupling \cite{Fan:2026,Shen:2026,Wang:2026,SunWJ:2026,Miao:2026}. However, the superconducting state in bulk single-crystalline \LNO is stabilized only under high pressure, a condition that severely restricts direct access by surface-sensitive and momentum-resolved probes such as STM/STS and ARPES. High-pressure point-contact Andreev-reflection spectroscopy (PCAS) have provided valuable evidence for the superconducting gap, but these techniques yield limited momentum information and have led to different interpretations, including two-gap $s$-like spectra \cite{Guo:2025,LiuPCAS:2025} and signatures of $d$-wave pairing \cite{Cao:2025}. A direct spectroscopic determination of the gap structure in pressurized single crystals is still lacking.



The normal state from which superconductivity emerges is equally unresolved. Rather than hosting a single parent instability, \LNO exhibits multiple density-wave-like anomalies with distinct pressure dependences at 130$\sim$150~K \cite{Sreedhar:1994,Taniguchi:1995,LING:2000517,Liu2022,Zhangjunjie:2023,WutaoNMR:2025,RIXS:2024,Shulei:2024,Kakoi:2024,Khasanov:2025}. One density-wave order is rapidly suppressed at low pressures below about 3 GPa, whereas another shows an enhanced transition temperature or 
strengthened magnetic character upon compression \cite{WutaoNMR:2025,Khasanov:2025}. These observations have been interpreted in terms of intertwined spin-density-wave (SDW) and charge-density-wave (CDW) \cite{ChenReview:2026}, but their hierarchy and relevance to superconductivity remain unclear. Importantly, such complexity means that the electronic order proximate to superconductivity need not be a simple high-pressure continuation of the ambient-pressure density-wave state. A probe capable of accessing low-energy electronic excitations directly under pressure, while retaining sensitivity to momentum-space anisotropy, is therefore required to identify the actual competing state adjacent to superconductivity. Polarization-resolved electronic Raman scattering is well suited to this task: through symmetry-dependent Raman vertices, it selectively weights different regions of the Brillouin zone (BZ) \cite{Devereaux:2007} and has already been shown to resolve anisotropic electronic correlations and momentum-selective SDW gaps in single-crystalline \LNO at ambient pressure \cite{He:2026}. These unresolved issues motivate a pressure-dependent, symmetry-resolved Raman study of single-crystalline \LNO to identify the electronic instability proximate to superconductivity and determine the superconducting gap structure that constrains the pairing symmetry.

Here we use polarization-resolved electronic Raman spectroscopy under high pressure to address both questions in single-crystalline \LNO. We identify a hidden order collective mode within the dominant stripe-ordered SDW state, which emerges below approximately 130 K between 9 and 14 GPa and may originate from fluctuations associated with a subleading double-$\boldsymbol{Q}$ magnetic/charge state, which appears in proximity to the structural transition. This mode is progressively suppressed at higher pressures and disappears as superconductivity emerges. In the superconducting state, pair-breaking peaks appearing in the \Blg and \BZg symmetries enable a quantitative determination of the superconducting gap anisotropy. Our results reveal weakly anisotropic s-wave superconducting gap on mutiple 
Fermi-surface pockets, establish a close interplay between the collective mode
and superconductivity, and provide direct spectroscopic constraints on the pairing state of bilayer nickelates.

\section{RESULTS}

\subsection{High-pressure Raman experiments} 

To establish a comprehensive pressure-temperature evolution of the electronic states in \LNO, we performed systematic polarization-resolved Raman scattering measurements under high pressure using a diamond-anvil cell (DAC) (Figs.~\ref{fig:rawdata}(a) and \ref{fig:rawdata}(b); see Appendix~A and Sec.~A of the Supplemental Material for details). To analyse the electronic Raman response, we use the $D_{4h}$ point-group notation. Although the crystal structure has been reported to evolve from $Amam$ to $Fmmm$ and eventually to $I4/mmm$ with increasing pressure \cite{Wang:2024,Sun:2023,WangNN:2024}, the associated lattice distortions remain small, making $D_{4h}$ a good approximation for classifying the in-plane electronic Raman response, as discussed in ref. \cite{He:2026}. Within this notation, the Raman response in the $ab$ plane can be decomposed into the \Alg, \Blg and \BZg symmetry channels using four basic polarization configurations, $xx$, $xy$, $x'x'$ and $x'y'$ (Fig.~\ref{fig:rawdata}(c); see Sec.~B of the Supplemental Material for details).

Figures~\ref{fig:rawdata}(d) and \ref{fig:rawdata}(e) show representative low-temperature Raman spectra measured at 3 K under 11.5 GPa and 19.5 GPa, respectively. In addition to sharp Raman-active phonon modes, the spectra exhibit pronounced electronic features. At 11.5 GPa, we observe an excitation near 177~\wn, marked by stars in Fig.~\ref{fig:rawdata}(d), together with a higher-energy feature around 800~\wn, marked by triangles. The latter corresponds to the SDW response identified in our previous Raman study \cite{He:2026}, whereas the former represents a distinct hidden order excitation that emerges only under pressure. Because its microscopic origin cannot be assigned unambiguously at this stage, we refer to this pressure-induced state as a hidden order (HO). At 19.5 GPa, where superconductivity is established, a new low-energy electronic response appears near 200~\wn, marked by diamonds in Fig.~\ref{fig:rawdata}(e). Although the HO and the superconducting response occur on a similar energy scale, their polarization dependence and temperature evolution are markedly different, indicating that they originate from distinct electronic states. The full pressure- and temperature-dependent evolution of these responses reveals how SDW correlations, HO and superconductivity appear and disappear across the phase diagram, as analysed in detail below.



\subsection{Identification of the hidden order (HO) collective mode} 
\begin{figure}[ht]
\centering
   \includegraphics[width=\textwidth]{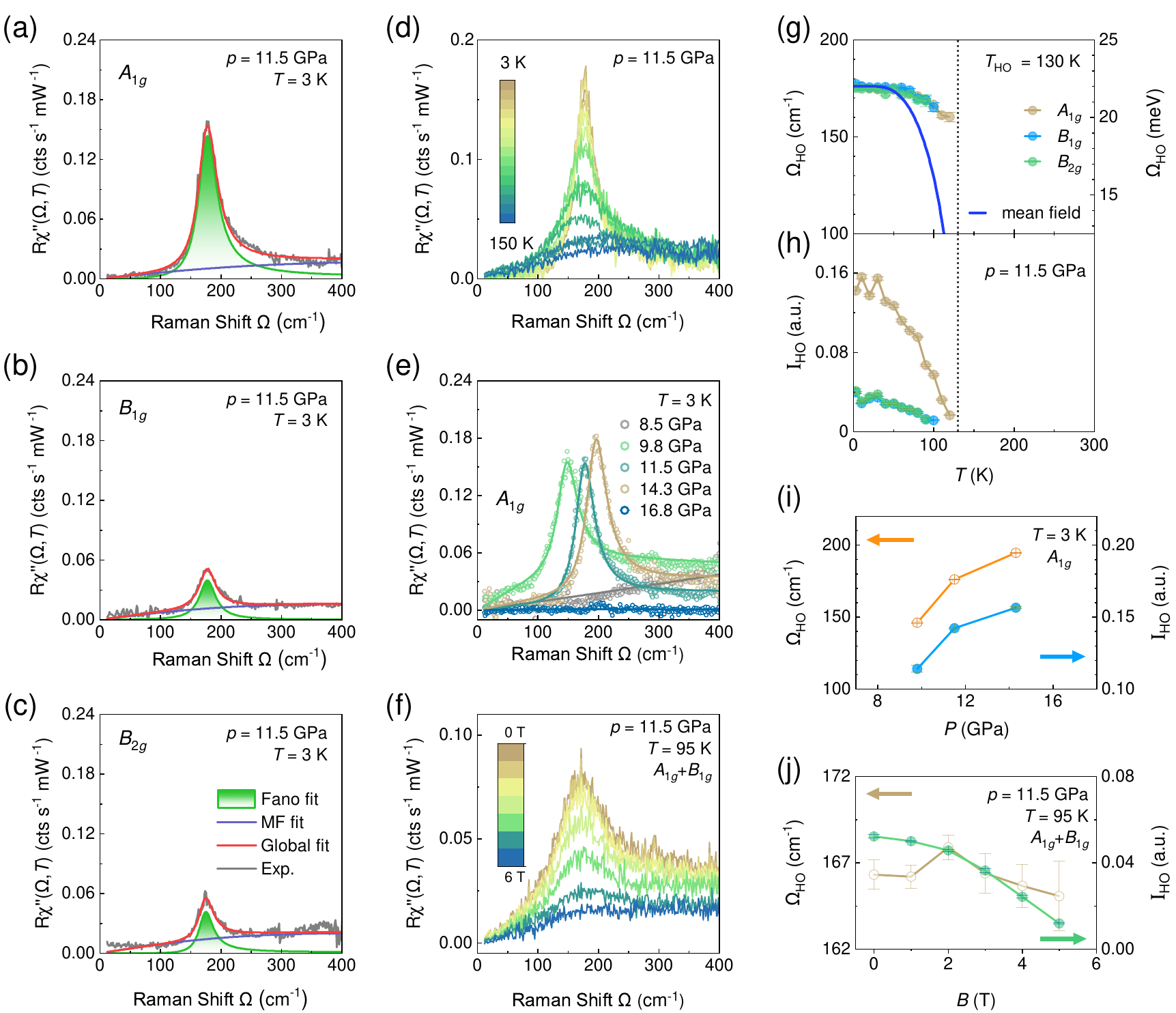}
   \caption{
Temperature, pressure, and polarization dependence of the hidden order collective mode excitation in \LNO.
(a)--(c) Fits to the electronic continuum in the \Alg, \Blg and \BZg channels at 3\,K and 11.5\,GPa using phenomenological memory-function (MF)-Fano model. 
(d) \Alg Raman spectra measured at selected temperatures.
(e) \Alg\ Raman response measured at 3\,K under various pressures.
(f) \Alg+\Blg Raman response measured at 95\,K under magnetic fields up to 6\,T.
(g),(h) Temperature dependence of the characteristic collective mode energy $\Omega_{\mathrm{HO}}$ and the integrated spectral weight $I_{\mathrm{HO}}$ attributed to the pure HO state, respectively.
(i) Pressure dependence of $\Omega_{\mathrm{HO}}$ and the integrated spectral weight $I_{\mathrm{HO}}$ at 3\,K.
(j) Magnetic-field dependence of $\Omega_{\mathrm{HO}}$ and the integrated spectral weight $I_{\mathrm{HO}}$ at 95\,K.
}
   \label{fig:HO}
\end{figure}

Between 9 and 14 GPa, we observe an additional electronic excitation near 177\,\wn, which has no counterpart in the ambient-pressure SDW response and is distinct from the superconducting pair-breaking feature observed at higher pressure. To quantify this excitation, we analyse the Raman spectra at 11.5 GPa and 3 K in the pure symmetry channels, \Alg, \Blg and \BZg (Figs.~\ref{fig:HO}(a)--\ref{fig:HO}(c)). The \Alg response is approximately three times stronger than the \Blg and \BZg responses, and we have confirmed that the weak signals in the latter channels do not originate from leakage of the \Alg component (see Sec.~C of the Supplemental Material for details). The electronic continuum is described using a memory-function (MF) form \cite{Gotze:1972,Opel:2000,Sen:2020}, whereas the HO near 177\,\wn is modeled by a Fano profile due to the asymmetric lineshape (see Sec.~D of the Supplemental Material for details). It is worth noting that the Fano lineshape applied in this work is intended purely as a phenomenological description of the Raman response, enabling extraction of the excitation energy and intensity, and carries no implication of a genuine discrete bosonic mode coupled to the electronic continuum. This procedure separates the HO contribution from the broad electronic background and allows us to extract its characteristic energy, $\Omega_{\mathrm{HO}}$, and integrated spectral weight, $I_{\mathrm{HO}}$, as a function of temperature and pressure.

The temperature dependence provides direct evidence for a collective mode. As shown in Figs.~\ref{fig:HO}(d), \ref{fig:HO}(g), and \ref{fig:HO}(h), $I_{\mathrm{HO}}$ is progressively suppressed upon warming and vanishes at $T_{\mathrm{HO}} = 130 \pm 5$ K at 11.5 GPa. Over the same temperature range, $\Omega_{\mathrm{HO}}$ softens from 177\,\wn at low temperature to 160\,\wn near $T_{\mathrm{HO}}$. This behavior cannot be captured by a simple mean-field order-parameter form (blue curve in Fig.~\ref{fig:HO}(g)). Moreover, the ratio $\Omega_{\mathrm{HO}}/k_{\mathrm{B}}T_{\mathrm{HO}} \sim 1.9$ is substantially smaller than the weak-coupling mean-field value of 3.53, suggesting that the HO excitation is not a conventional density-wave gap with a simple mean-field temperature evolution.

We next track the pressure evolution of the HO response. In the \Alg channel, the peak associated with the HO phase becomes visible above about 9 GPa and is no longer observed above about 16 GPa (Figs.~\ref{fig:HO}(e) and \ref{fig:HO}(i)). In this pressure range, its spectral weight shows a clear positive pressure dependence, at the same time, $\Omega_{\mathrm{HO}}$ increases from 146 to 195\,\wn (Figs.~\ref{fig:HO}(e) and \ref{fig:HO}(i)). This pressure evolution indicates that the HO is stabilized only in an intermediate-pressure regime.

To further explore the origin of the HO, we examine the magnetic-field response of the HO excitation. Figures~\ref{fig:HO}(f) and \ref{fig:HO}(j) show field-dependent Raman spectra measured in \Alg + \Blg channel at 95 K up to 6 T. The HO peak is gradually suppressed with increasing magnetic field and disappears at 6 T, whereas its characteristic energy changes only weakly. The strong field sensitivity of the spectral weight points to a magnetic component or magnetic coupling associated with the HO. These results establish the HO as a distinct pressure-induced electronic collective mode and motivate the comparison with superconductivity in the following section.



\subsection{Superconducting gap structure} 
\begin{figure}[ht]
\centering
   \includegraphics[width=\textwidth]{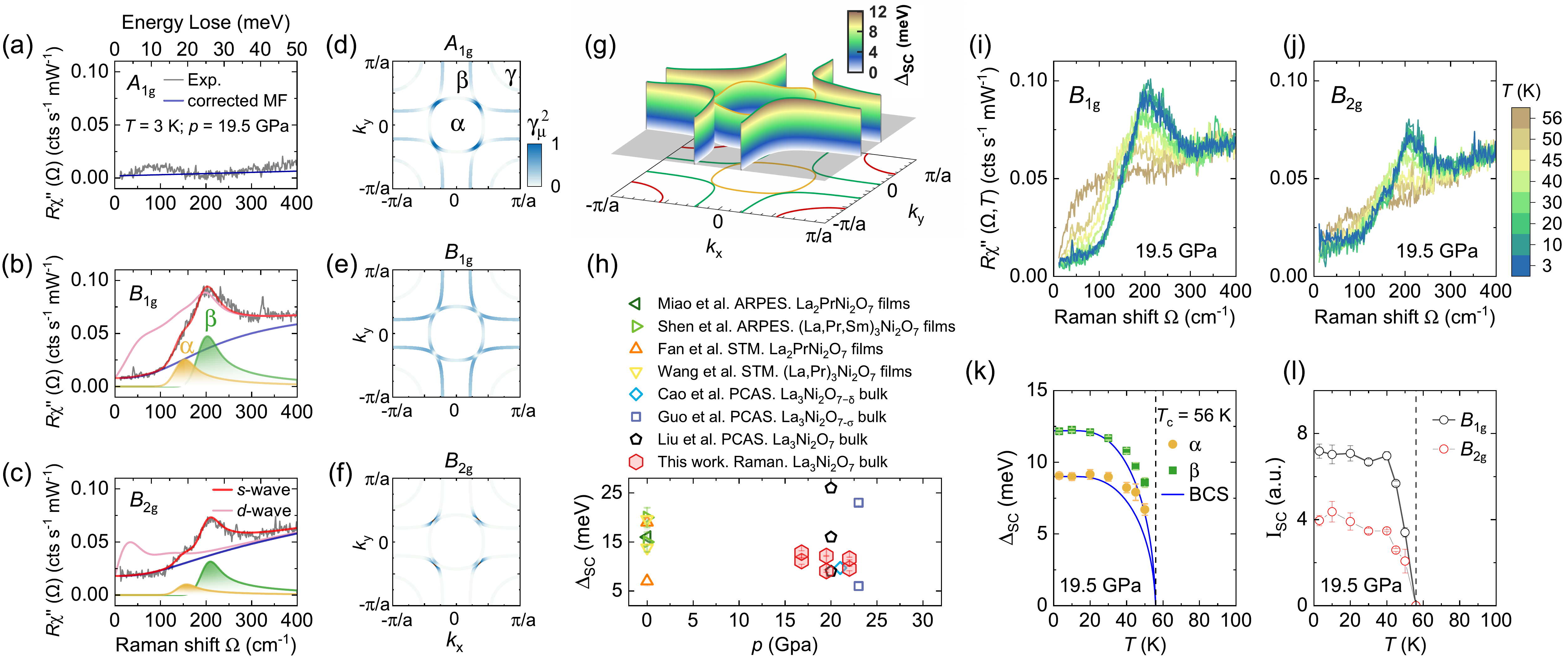}
   \caption{
Superconducting gap structure and symmetry analysis in pressurized \LNO.
(a)--(c) Quantitative fits to the superconducting Raman response in the \Alg, \Blg, and \BZg\ symmetry channels, respectively. The corrected memory-function (MF) model is used to describe the normal-state electronic scattering (blue curves), which is subtracted prior to fitting the superconducting response. An anisotropic $s$-wave (red curve) and a $d$-wave (pink curve) superconducting gap function are used to fit the experimental data.
(d)--(f) Corresponding Raman vertices calculated within the effective-mass approximation.
(g) Momentum-dependent superconducting gap obtained from the symmetry-resolved Raman analysis.
(h) Comparison of superconducting gap values for Ruddlesden--Popper bilayer nickelate thin films and single crystals reported by STM \cite{Fan:2026,Wang:2026}, ARPES \cite{Shen:2026,Miao:2026}, point-contact spectroscopy \cite{Guo:2025,Cao:2025,LiuPCAS:2025}, and Raman spectroscopy (this work).
(i)--(j) Temperature evolution of the superconducting Raman response in the \Blg\ and \BZg\ symmetry channels, respectively.
(k)--(l) Temperature dependence of the superconducting pair-breaking peak energy $\Delta_{\mathrm{SC}}$ and the integrated spectral weight of the pure superconducting response, $I_{\mathrm{SC}}$, in the \Blg\ and \BZg\ symmetry channels, respectively. The gap energies exhibit a temperature dependence that deviates slightly from the prediction of BCS theory (blue curves in (k)).
}
   \label{fig:SC}
\end{figure}

We now turn to the superconducting state at 19.5 GPa. Figures~\ref{fig:SC}(a)--\ref{fig:SC}(c) show the superconducting Raman response in the \Alg, \Blg and \BZg symmetry channels at 3\,K. In the superconducting state, pronounced low-energy pair-breaking features are observed in the \Blg and \BZg  channels. By contrast, the \Alg response is extremely weak, preventing a reliable identification of an analysable pair-breaking signal, probably owing to strong Coulomb screening and/or multiband effects \cite{Monien:1990,Lazarevic:2020,Devereaux:2007}. The \Blg response develops a broad feature that onsets near 100\,\wn and peaks at approximately 200\,\wn. The \BZg response also onsets near 100\,\wn, shows a kink-like feature around 160\,\wn, and reaches its maximum at a slightly higher energy of about 210\,\wn. This pronounced channel dependence indicates that the Raman response is not governed by a single isotropic gap scale, but instead reflects a momentum-dependent superconducting gap sampled differently by the symmetry-selective Raman vertices. 

To determine the superconducting gap quantitatively, we fit the spectra using a two-step procedure. The normal-state electronic continuum is first described by a depletion-corrected MF model and subtracted as the background (see Sec.~D of the Supplemental Material for details). The remaining superconducting response is then analysed using the Tsuneto-Maki (TM) Raman response calculated from a tight-binding band structure (see Secs.~E and F of the Supplemental Material for details). We use the usual convention for the Fermi pockets, i.e. $\alpha$ pocket circling around the $\Gamma$ point and $\beta$ and $\gamma$ pockets circling around the M point. The superconducting Raman response at $T=0$ is given by \cite{Tsuneto:1960,Bohm:2014}

\begin{equation}
\label{eq:SCRaman}
\chi_{\mu}{''}(\Omega)=4\pi\sum_{n}\left\langle \frac{\gamma^2_{n,\mu}(\mathbf{k})\left|2\Delta_n(\mathbf{k})\right|^2}{\Omega\sqrt{\Omega^2-\left|2\Delta_n(\mathbf{k})\right|^2}}\right\rangle,
\end{equation}

\noindent where $\mu$ denotes the Raman symmetry channel (\Alg, \Blg, or \BZg), $\gamma_{n,\mu}(\mathbf{k})$ is the Raman vertex for band $n$ at momentum $\mathbf{k}$ in symmetry channel $\mu$, $\Delta_n(\mathbf{k})$ is the corresponding superconducting gap function, the angled brackets denote an average over the Fermi surface of band $n$, and $\Omega>\left|2\Delta_n(\mathbf{k})\right|$.
The Raman vertices for the three symmetry channels are estimated using effective mass approximation \cite{Devereaux:2007} as shown in  Figs.~\ref{fig:SC}(d)--\ref{fig:SC}(f). As we can see, \Alg strongly weights the $\alpha$ pocket along the diagonal direction, and anisotropic weights the $\beta$ pocket with stronger intensity near the zone boundary. \Blg channel mainly projects out the excitation from $\alpha$ and $\beta$ pockets away from the diagonal direction. \BZg channel mainly projects out the diagonal region of the BZ for both $\alpha$ and $\beta$ pockets. Here, the Raman vertex only describes the electron-photon coupling strength; the measured Raman intensity is further modulated by the Fermi-surface density of states. In particular, the \Alg response should also be considered together with the screening effect \cite{Monien:1990,Devereaux:2007}. We test anisotropic $s$-wave ($\Delta(\mathbf{k}) = \Delta_0\frac{1+r\cos\left[4(\theta+\theta_0)\right]}{1+r}$) and $d$-wave ($\Delta(\mathbf{k})=\Delta_0\cos(2\theta)$) gap functions, and perform a simultaneous fit to the \Blg and \BZg spectra using one common set of superconducting gap parameters.

The fits directly give the gap size and symmetry. The best fitting constrain the maximum gap at $\alpha$ and $\beta$ pocket are $\Delta_{\alpha} = 9.0$ meV and $\Delta_{\beta} = 12.2$ meV at 19.5 GPa, corresponding to gap ratios of  2$\Delta_{\alpha}/k_B T_{\mathrm{c}} \approx$ 3.7 and 2$\Delta_{\beta}/k_B T_{\mathrm{c}} \approx$ 5.1, respectively. This global symmetry-resolved analysis strongly constrains the possible gap symmetry. The $d$-wave form produces excessive low-energy spectral weight and different positions of the pair-breaking peaks in the \Blg and \BZg symmetries that is incompatible with our observed Raman response (pink curves in Figs.~\ref{fig:SC}(b) and \ref{fig:SC}(c)). By contrast, an anisotropic $s$-wave gap consistently captures the peak positions, lineshapes and relative intensities in both \Blg and \BZg symmetry channels. The anisotropy for $\alpha$ and $\beta$ pockets is approximately r = 0.05. The resulting momentum-dependent gap distribution is shown in Fig.~\ref{fig:SC}(g), revealing multiple fully gapped Fermi-surface pockets with weak gap anisotropy. Note that we are unable to determine the gap structure on $\gamma$ pocket due to the weak Raman vertex in the \Blg and \BZg symmetries.

Figure~\ref{fig:SC}(h) plots the pressure dependent gap size.  The gap shows weak pressure dependence from 16.5 to 21 GPa. We then compare the energy scales obtained here with those reported by other spectroscopic probes of RP bilayer nickelates (Fig.~\ref{fig:SC}(h)), including STS \cite{Fan:2026,Wang:2026}, ARPES \cite{Miao:2026,Shen:2026} and PCAS \cite{Guo:2025,Cao:2025,LiuPCAS:2025}. These measurements have reported multiple superconducting gap scales in both thin films and pressurized single crystals. The smaller gap is compatible,  but the larger gap obtained from thin films and PCAS in bulk samples is larger than those obtained from our Raman experiments. These discrpancy may own to difference in hydrostatic pressure in bulk but in-plane pressure stress in thin film  and/or directly probe dependence. Despite differences in sample form and experimental environment, these measurements broadly converge on a multigap superconducting state with weak anisotropy and no symmetry-enforced nodes. Our Raman results provide a high-pressure and symmetry-resolved determination of this gap structure in \LNO bulk samples, supporting an anisotropic $s$-wave pairing state in pressurized \LNO.

We further examine the temperature dependence of the superconducting Raman response. As shown in Figs.~\ref{fig:SC}(i) and \ref{fig:SC}(j), the pair-breaking peaks in both the \Blg and \BZg channels are progressively suppressed upon warming and shift markedly to lower energy as the temperature approaches $T_{\mathrm{c}}$. This softening is summarized in Fig.~\ref{fig:SC}(k), where $\Delta_{\mathrm{SC}}$ decreases continuously with increasing temperature, with a slight deviation from the conventional BCS temperature dependence. Meanwhile, the integrated superconducting spectral weight, $I_{\mathrm{SC}}$, is gradually depleted and vanishes near $T_{\mathrm{c}}$ (Fig.~\ref{fig:SC}(l)). Above $T_{\mathrm{c}}$, closure of the superconducting gap reopens the low-energy quasiparticle scattering channel, leading to a pronounced increase in the initial low-energy slope of the Raman response, particularly in the \Blg channel (Figs.~\ref{fig:SC}(i) and \ref{fig:SC}(j)). These behaviours are characteristic Raman signatures of superconducting gap opening and closing, as observed in Pr$_{2-x}$Ce$_x$CuO$_4$ \cite{Qazilbash:2005}, Bi-2212 \cite{Sugai:2003} and Ba$_{0.6}$K$_{0.4}$Fe$_2$As$_2$ \cite{Bohm:2014}.

Under magnetic field, the intensities of the pair-breaking peaks in the \Blg and \BZg channels are suppressed by nearly 50\% up to 6\,T, whereas their peak energies remain nearly unchanged. This behaviour can be understood as a consequence of field-induced low-energy quasiparticles generation and superconducting coherence suppression. At the same time, the large upper critical field of \LNO prevents a significant suppression of the gap energy under the relatively low fields applied here.

\subsection{Pressure-dependent phase diagram} 
\begin{figure}[ht]
\centering   \includegraphics[width=0.8\textwidth]{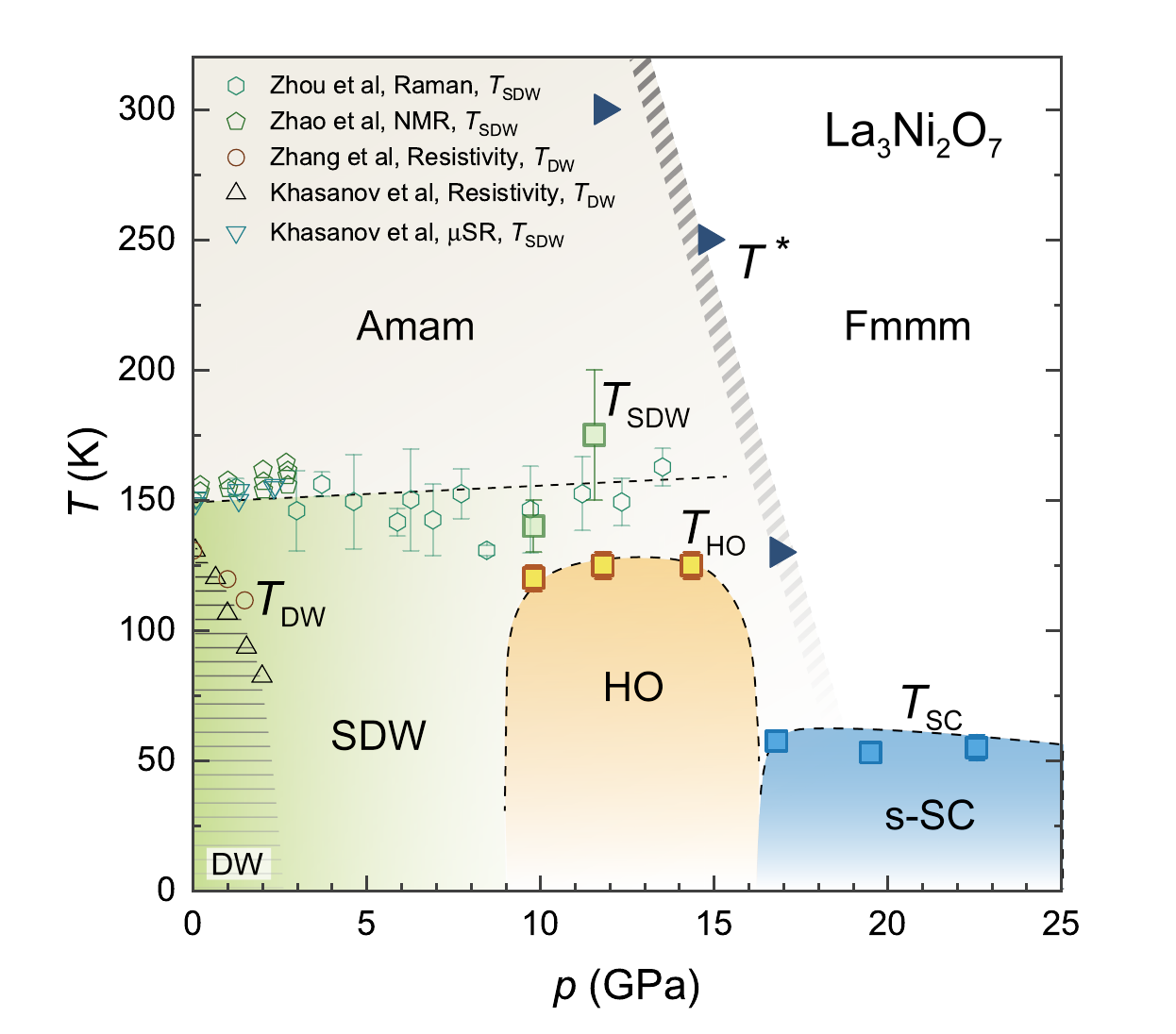}
   \caption{
Pressure-dependent electronic phase diagram of \LNO.
The spin-density-wave transition temperature $T_{\mathrm{SDW}}$ and the density-wave transition temperature $T_{\mathrm{DW}}$ are adopted from Refs.~\cite{Zhang:2024,Khasanov:2025,WutaoNMR:2025,Zhou:2026}. Two data points of $T_{\mathrm{SDW}}$ are obtained from this work (light green squares). The structural transition temperature $T^{*}$ (blue triangles) is determined from anomalies in the Raman-active phonons associated with the structural transition. The HO collective mode transition temperature $T_{\mathrm{HO}}$ (yellow squares) is extracted from the temperature dependence of the integrated spectral weight of the \Alg\ peak. The superconducting transition temperature $T_{\mathrm{SC}}$ is determined from the temperature dependence of the pair-breaking peak spectral weight in the \Blg\ and \BZg\ Raman spectra.
}
   \label{fig:phasedia}
\end{figure}

Based on the systematic pressure- and temperature-dependent Raman dataset, we construct an electronic pressure-temperature phase diagram of \LNO, as shown in Fig.~\ref{fig:phasedia}. At ambient pressure, previous work has revealed two density-wave orders \cite{Khasanov:2025}, which we label as DW and SDW. The origin of the DW phase is still unknown, presumably associated with CDW. Upon compression, DW is rapidly suppressed below 3 GPa, while the transition temperature of SDW increases and persists up to about 14 GPa \cite{Zhou:2026}, as reported in our previous Raman results. Note that the enhancement of SDW under pressure up to 13 GPa has also been reported in La$_5$Ni$_3$O$_{11}$ compounds \cite{Shi:2025}.

The new phase identified in this work is a HO phase, which collective mode emerges between 9 and 14 GPa below approximately 130 K. As we show below it may be associated with a subleading double-$\boldsymbol{Q}$ magnetic/charge instability, which appears close to the orthorhombic/tetragonal transition. The structural transition from $Amam$ to $Fmmm$, which has approximate $C_4$ symmetry of the low-energy electronic structure, occurs between 14.3 and 19.5 GPa, as confirmed by the pressure evolution of the phonon modes \cite{ZhangHY:2026}. The superconducting phase boundary is determined from the disappearance of the superconducting pair-breaking response upon warming. The superconducting transition temperature, $T_{\mathrm{c}}$, is approximately 55 K, comparable to the zero-resistance transition observed in electrical transport measurements \cite{Zhang:2024}. Within the measured pressure range, $T_{\mathrm{c}}$ exhibits only a weak pressure dependence, consistent with previous reports \cite{Sun:2023}.
 
\section{Discussion and Conclusion}

The HO collective mode identified here is a distinct pressure-induced excitation in \LNO. In contrast to the SDW response, which appears at a higher energy scale and shows different symmetry, temperature, magnetic-field, and pressure dependences \cite{Khasanov:2025,WutaoNMR:2025,He:2026,RIXS:2024,KeeHoonKim:2026}, the HO mode emerges only above approximately 9 GPa, is confined to an intermediate-pressure regime preceding robust superconductivity, and is suppressed at higher pressures. Its pronounced magnetic-field sensitivity, together with its weak energy renormalization, suggests that it is not an amplitude mode of the established stripe-ordered SDW state with the single $Q_y=(0,\pi,\pi)$ ordering wave vector. Instead, it may represent a collective excitation associated with a competing double-$\boldsymbol{Q}$ magnetic instability with both $Q_x=(\pi,0,\pi)$ and $Q_y=(0,\pi,\pi)$ ordering  wavevectors and  that becomes increasingly favorable upon approaching the pressure-induced orthorhombic-to-tetragonal structural transition. Notably, recent high-pressure transport measurements have identified a third characteristic temperature, $T_3$, in the 10-14 GPa range \cite{Kriener:2026}, which may share a common origin with the HO collective mode observed here.

To verify this prediction we extended the Hartree-Fock calculations of Ref. \cite{Braz:2026} starting not from $Amam$ phase but from tetragonal $I4/mmm$ phase where we adopted the DFT parameters from Ref. \cite{Jiang:2025} for various pressures. The resulting phase diagram is shown in Fig.~S8 of the Sec.~G of the Supplemental Material. In contrast to the $Amam$ phase where the double stripe spin-charge state appears to be always dominant, in the tetragonal phase there are several other double-$\boldsymbol{Q}$ phases apart from double spin-charge stripe phase, which also breaks the tetragonal symmetry. Most notable is tilted vortex phase (TVP), which is characterized by two non-equivalent magnetizatitons $M_{Q_x} < M_{Q_y}$, which appears around the same pressure region. Such a state is only stable if one performs the calculations in the tetragonal phase but it becomes subleading instability once the calculations are done for the $Amam$ phase. In other words the TVP phase never represents the true ground state but is close in energy. Note that TVP phase is also accompanied by the checkerboard charge order, which means that it should have a corresponding coupled spin-charge collective modes.

A natural interpretation is therefore that the HO collective mode originates from fluctuations toward this subleading double-$\boldsymbol{Q}$ magnetic TVP state competing with the dominant single-$\boldsymbol{Q}$ double stripe-charge ordered SDW state. Such a double-$\boldsymbol{Q}$ state is expected to become more competitive as the orthorhombic distortion is reduced and tetragonal symmetry is approached under pressure. At a phenomenological level, this situation is analogous to the Bardasis–Schrieffer modes observed in iron-based superconductors \cite{Bohm:2014,He:2020}, in which a collective excitation develops in a subdominant ordering channel competing with the dominant ground-state order. Here, however, the observed mode is not a superconducting Bardasis–Schrieffer excitation, but rather a particle–hole channel collective mode associated with a subleading double-$\boldsymbol{Q}$ magnetic order. Because Raman scattering primarily couples to charge excitations, the observation of this mode implies that the incipient double-$\boldsymbol{Q}$ magnetic instability is accompanied by, or strongly coupled to, a secondary charge component. 
This is indeed the case for the TVP as shown in the Sec.~G of the Supplemental Material. Such spin–charge coupling naturally accounts for the Raman activity of the HO mode and its strong sensitivity to magnetic field. Its polarization dependence provides further support for this competing-channel interpretation. Whereas the stripe-related SDW ground state is most prominent in the \Blg and \BZg channels, the HO collective mode appears predominantly in the nearly orthogonal \Alg channel, consistent with a collective excitation of a competing double-$\boldsymbol{Q}$ instability favored near the tetragonal phase, rather than an amplitude mode of the single-$\boldsymbol{Q}$ stripe order. The substantially weaker HO signals in the \Blg and \BZg channels remain unresolved and may arise from residual symmetry mixing or coupling among the competing magnetic and charge channels.

The superconducting Raman response places direct constraints on the gap structure of pressurized bulk \LNO. Although the \Alg channel does not yield a reliable pair-breaking signal, the \Blg and \BZg responses provide sufficient symmetry selectivity to resolve the gaps on the $\alpha$ and $\beta$ pockets using a common set of fitting parameters. The extracted gap structure reveals multiband superconductivity with a fully opened but anisotropic gap, broadly consistent with the nodeless multigap behaviour reported in La$_3$Ni$_2$O$_7$ thin films \cite{Shen:2026,Miao:2026,Fan:2026,Wang:2026}. Thus, our Raman measurements establish, independently and under high pressure, multiband superconductivity in single-crystalline \LNO. The same analysis disfavors a $d$-wave gap and is instead naturally described by an anisotropic $s$-wave state, consistent with theoretical scenarios in which interlayer pairing involving Ni $d_{z^2}$ states gives rise to a sign-changing $s_{\pm}$-like gap structure \cite{WangReview:2024,Liu:2023,Ilya:2024,Qu:2024,ZhangNC:2024,Zhan：2026}. An important open question is the superconducting gap on the $\gamma$ pocket, which cannot be directly resolved in the present experiment because of the weak \Alg response. Future high-sensitivity Raman measurements may determine whether the $\gamma$ pocket is fully gapped and, if so, how its gap anisotropy fits into the multiband pairing structure.

The relationship between ordered states and superconductivity in \LNO is more complex than a simple density-wave-to-superconductor transition. Previous studies have revealed several competing electronic instabilities, including a low-pressure density-wave state and an SDW order that persists close to the superconducting regime. Our observation of the HO collective mode further points to an additional competing tetragonal magnetic instability in the same pressure range. However, it remains unclear whether superconductivity microscopically coexists with any of these ordered states or replaces them locally. As a result, no single ordered phase can yet be assigned unambiguously as the sole parent of superconductivity in \LNO. This view is consistent with theoretical work predicting a pressure-dependent competition among stripe-type magnetism \cite{RIXS:2024,Shulei:2024,Ni:2025,ZhangNC:2024,Braz:2026}, density waves \cite{WangYL:2024} and interlayer-exchange-driven instabilities involving Ni $d_{z^2}$ orbitals \cite{ChenReview:2026,Qu:2024}. In this context, the Raman-observed HO collective mode provides spectroscopic evidence for a proximate competing magnetic instability, adding an important constraint on the route to superconductivity under pressure.

In summary, our high-pressure Raman measurements reveal a pressure-driven succession from competing magnetic states to multiband superconductivity in single-crystalline \LNO. The emergence and subsequent disappearance of a field-sensitive collective mode near the superconducting regime, together with the observation of nodeless, weakly anisotropic $s$-wave gaps, provide key spectroscopic constraints on both the parent electronic states and the pairing symmetry. These results suggest that fluctuations generated by the competition and suppression of magnetic orders may contribute to superconducting pairing. Direct determination of the high-pressure magnetic structure and fluctuation spectrum, clarification of the microscopic relationship between magnetism and superconductivity, and phase-sensitive measurements of the complete gap structure will be essential for establishing whether this interplay represents a common organizing principle of unconventional superconductivity.

\section*{Acknowledgments}

This work was supported by the National Natural Science Foundation of China (Grant Nos. 12425404, 12474473, U25A20193, 12494591, 12474137, and 92565303), the National Key Research and Development Program of China (Grants No. 2023YFA1406000, 2023YFA1406500, and 2024YFF0727103), the Fundamental and Interdisciplinary Disciplines Breakthrough Plan of the Ministry of Education of China (JYB2025XDXM403), the Guangdong Major Project of Basic Research (2025B0303000004), the Guangdong Basic and Applied Basic Research Funds (Grant No. 2024B1515020040, 2025B1515020008, 2024A1515030030), the Shenzhen Science and Technology Program (Grants No. RCYX20231211090245050), the Guangzhou Basic and Applied Basic Research Funds (Grant No. 2024A04J6417), the CAS Superconducting Research Project (Grant No. SCZX-0101), the Guangdong Provincial Key Laboratory of Magnetoelectric Physics and Devices (Grant No. 2022B1212010008), and the Research Center for Magnetoelectric Physics of Guangdong Province (Grant No. 2024B0303390001), L.B.B. acknowledges financial support from the grants 2023/14902-8 and
2025/17852-7, Sao Paulo Research Foundation (FAPESP).\\ 


M.W. conceived the project. H.Y.Z, H.L.S., G.H., J.C. and B.S. performed the Raman measurements. D.Y.H. and C.H. synthesized and characterized the samples. M.W., G.H., I.E., J.Shen, H.Y.Z., L.B. and J.Shu analyzed the data. G.H., M.W., J.Shen, H.L.S., J.Shu and H.Y.Z. wrote the manuscript with comments from all the authors. \\

\noindent\textbf{Conflicts of Interest} 

The authors declare no competing interests.\\ 

\noindent\textbf{Data Availability Statement} 

All relevant data that support the findings of this study are presented in the manuscript and supplementary information file. All data are available upon reasonable request from the corresponding authors.\\

\noindent\textbf{Methods}

\noindent\textbf{Samples}: Polycrystalline La$_3$Ni$_2$O$_7$ precursor rods were prepared by a conventional solid-state reaction. Pre-dried, high-purity La$_2$O$_3$ (99.99\%) and NiO (99.99\%) powders were weighed in stoichiometric proportions, thoroughly mixed and ground, and sintered at 1200\,$^\circ$C. Several grinding--sintering cycles were performed to improve compositional homogeneity. The resulting powder was isostatically pressed into cylindrical feed rods at 60\,MPa. The pressed rods, 5\,cm in diameter and 12\,cm in length, were subsequently sintered at 1400\,$^\circ$C for 24\,h to provide adequate mechanical strength for floating-zone growth. Single crystals were grown in a vertical optical floating-zone furnace (SciDre, model HKZ) equipped with a 5\,kW xenon arc lamp. Growth was carried out under flowing high-purity oxygen at 15\,bar and 0.15\,L\,min$^{-1}$. The feed and seed rods were counter-rotated, while the feed rod was translated at 3\,mm\,h$^{-1}$ to maintain a stable molten zone. The quality of the as-grown crystals was evaluated by Laue diffraction, X-ray diffraction and electrical-transport measurements, as detailed in Fig.~S1.

\noindent\textbf{Light scattering}: Raman scattering measurements were performed in a confocal backscattering geometry using a MoniVista CRS3 system coupled to a Princeton Instruments HRS-500 spectrograph. A 532\,nm laser was used for excitation, with incident power maintained below 3.5\,mW. Spectra were recorded using a 1,200 grooves\,mm$^{-1}$ grating, yielding a spectral resolution of 1.365\,cm$^{-1}$. A Zeiss 20$\times$ objective (NA 0.25) focused the excitation beam and collected the scattered light, which was detected using a nitrogen-cooled CCD. Single crystals of La$_3$Ni$_2$O$_7$ were loaded into a miniature Be--Cu diamond-anvil cell equipped with 250\,$\mu$m culets and a rhenium gasket. Helium served as the pressure-transmitting medium, providing quasi-hydrostatic conditions. The crystal $ab$ plane was aligned parallel to the diamond culets. The cell was mounted in an optical magnetic cryostat (Quantum Design) for \textit{in situ} measurements between 2 and 300\,K, and a Cernox sensor near the sample chamber monitored the temperature.

\bibliography{refs}
\newpage


\begin{figure}[ht]
\centering
   \includegraphics[width=\textwidth]{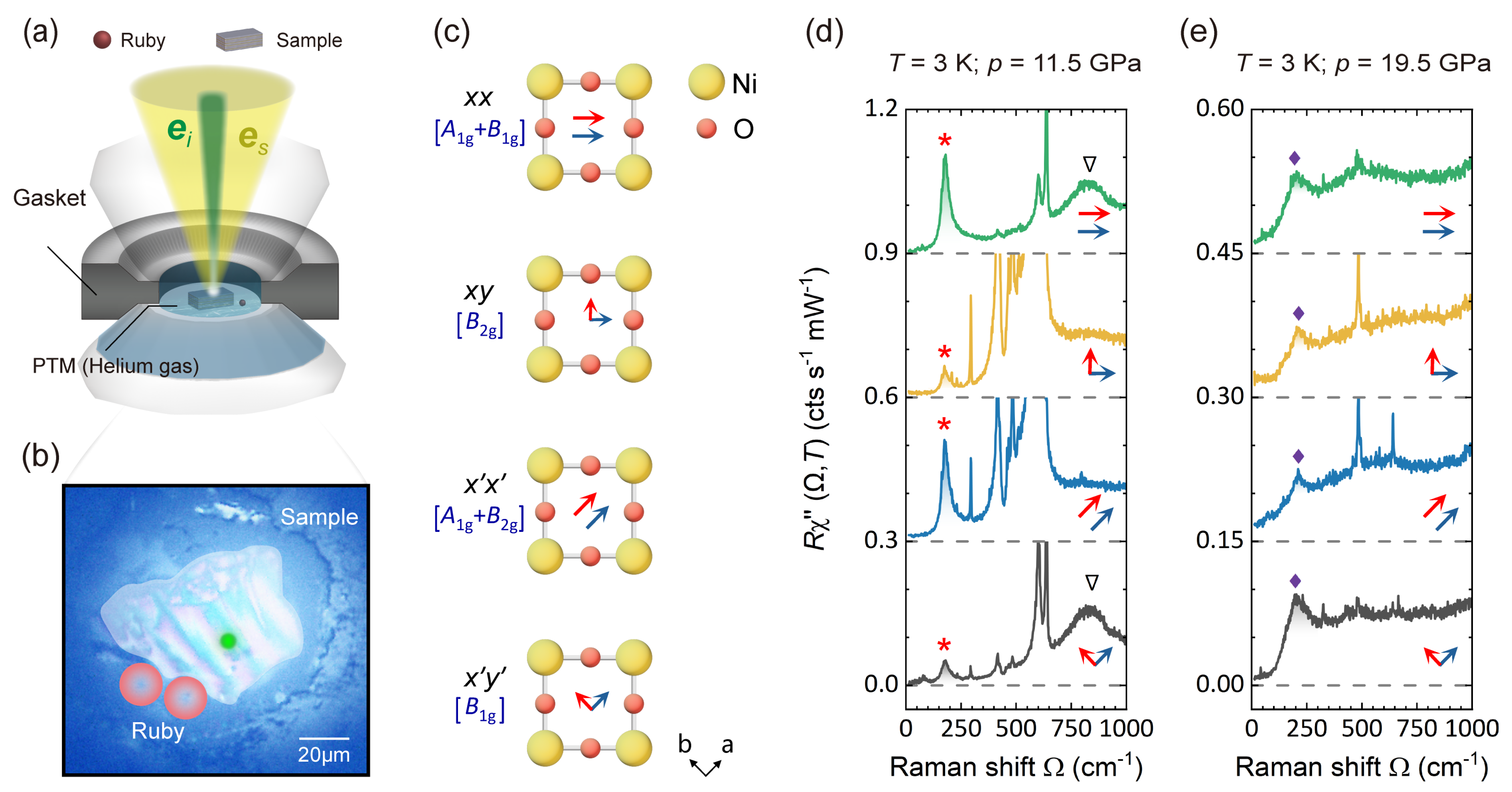}
   \caption{
High-pressure device, polarization configurations, and corresponding Raman spectra.
(a) Diamond anvil cell (DAC) with the sample sealed in rhenium using helium as the pressure-transmitting medium.
(b) Optical microscopy image of the sample inside the DAC. The green spot indicates the laser focus position.
(c) Schematic illustration of the polarization configurations. Red and yellow spheres denote oxygen and nickel atoms, respectively. The crystallographic $x$- and $y$-axes are aligned along the Ni-O-Ni bond directions, whereas the $x'$- and $y'$-axes are rotated by 45$^\circ$ clockwise with respect to the $x$ and $y$ axes. The corresponding Raman symmetry channels are defined within the $D_{4h}$ point group.
(d),(e) Raman spectra measured at 3\,K under pressures of 11.5\,GPa and 19.5\,GPa, respectively, for the corresponding polarization configurations. The electronic response associated with the collective excitation associated with a tetragonal magnetic instability is marked by stars, the SDW-related peak by triangles, and the superconducting pair-breaking peak by diamonds. The data other than those measured in the $x'y'$ channel are shifted vertically for clarity.}

   \label{fig:rawdata}
\end{figure}

\end{document}